\documentclass[conference]{IEEEtran} 

\usepackage{graphicx}

\usepackage[cmex10]{amsmath}
\usepackage{amssymb}                                                        
\usepackage{cite}
\usepackage{makeidx}
\usepackage{amsmath}
\usepackage{mathrsfs}
\usepackage{array}
\usepackage{tabularx}
\usepackage{multirow}
\usepackage{url}
\usepackage{tabularx}
\usepackage{booktabs}
\newtheorem{theorem}{Theorem}[section]

\newtheorem{remark}[theorem]{Remark}

\usepackage{mathptmx} 
\usepackage{times} 
\usepackage{amsmath} 
\usepackage{amssymb}  

\usepackage{algpseudocode}
\usepackage{algorithm}
\usepackage{tikz}
\usetikzlibrary{arrows}
\usepackage{tabularx}
\usepackage{dsfont}

\usepackage{caption}
\usepackage{subcaption}
\usepackage{dblfloatfix}


\makeatletter
\newcommand*\titleheader[1]{\gdef\@titleheader{#1}}
\AtBeginDocument{%
	\let\st@red@title\@title%
	\def\@title{%
		\bgroup\normalfont\large\centering\@titleheader\par\egroup
		\vskip0.5em\st@red@title}
}
\makeatother
\titleheader{2018 IEEE PES Innovative Smart Grid Technologies Conference, Washington D.C., USA}
\title{Battery Energy Storage Scheduling for Optimal Load Variance Minimization}
\IEEEoverridecommandlockouts                              

\begin{document}
\author{
	\IEEEauthorblockN{Yichen Zhang\IEEEauthorrefmark{1},
		Alexander Melin\IEEEauthorrefmark{2},
		Mohammed Olama\IEEEauthorrefmark{2},
		Seddik Djouadi\IEEEauthorrefmark{1},
		Jin Dong\IEEEauthorrefmark{2} and 
	    Kevin Tomsovic\IEEEauthorrefmark{1}}
	\IEEEauthorblockA{\IEEEauthorrefmark{1}Department of Electrical Engineering and Computer Science, University of Tennessee, Knoxville, TN 37996-2250}
	\IEEEauthorblockA{\IEEEauthorrefmark{2} Oak Ridge National Laboratory, Oak Ridge, TN 37831}
	\IEEEauthorblockA{Email: {\tt\small \{yzhan124, mdjouadi, tomsovic\}@utk.edu.}, {\tt\small \{melina, olamahussemm, dongj\}@ornl.gov.}}
	\thanks{Research sponsored by the Laboratory Directed Research and Development Program of Oak Ridge National Laboratory (ORNL), managed by UT-Battelle, LLC for the U.S. Department of Energy under Contract No. DE-AC05-00OR22725. The submitted manuscript has been authored by a contractor of the U.S. Government under Contract DE-AC05-00OR22725. Accordingly, the U.S. Government retains a nonexclusive, royalty-free license to publish or reproduce the published form of this contribution, or allow others to do so, for U.S. Government purposes.}}

\maketitle

\begin{abstract}
Generation portfolio can be significantly altered due to the deployment of distributed energy resources (DER) in distribution networks and the concept of microgrid. Generally, distribution networks can operate in a more resilient and economic fashion through proper coordination of DER. However, due to the partially uncontrollable and stochastic nature of some DER, the variance of net load of distribution systems increases, which raises the operational cost and complicates operation for transmission companies. This motivates peak shaving and valley filling using energy storage units deployed in distribution systems. This paper aims at theoretical formulation of optimal load variance minimization, where the infinity norm of net load is minimized. Then, the problem is reformulated equivalently as a linear program. A case study is performed with capacity-limited battery energy storage model and the simplified power flow model of a radial distribution network. The influence of capacity limit and deployment location are studied.

\end{abstract}

\section{INTRODUCTION}

Increasing deployment of renewable sources (mainly wind and solar photovoltaic (PV) generation) has decreased the power consumption for distribution companies. 
However, deep penetration of renewable resources into the current electric grid remains a challenging problem.
When there is a concurrent drop in renewable generation and increase in demand, the power consumption at the point of common coupling (PCC) between distribution networks and the main grid changes rapidly. This is known as the duck curve issue \cite{duck_curve}.
In particular, the most significant daily ramp starts around 5:00 p.m. when the sun sets (i.e., solar generation ends) and the demand increases \cite{sanandaji2016ramping}. 
The traditional power system is designed to meet the highest level of demand but ramping rates are limited particularly for large thermal units. Large amounts of renewable generation during off-peak hours deepen the valley and increase the ramp rate requirement. 
Fast ramping units, however, are limited in number and capacity, capital-intensive and subjected to possible transmission network congestions \cite{majzoobi2016application}. \par

An alternative solution is peak shaving and valley filling, referred to here as load variance minimization, by utilizing various dispatchable resources in distribution networks. Electric vehicles (EV) are scheduled in \cite{wang2013grid} so that the power at PCC follows a target profile. A storage scheduling algorithm that is resilient to the inevitable errors between the forecasted and actual demand is proposed in \cite{rowe2014peak} for peak demand reduction. A day-ahead battery energy storage scheduling is performed in low voltage unbalance distribution networks in \cite{joshi2015day}, where peak load shaving is considered as the main objective and load leveling is regarded as the second objective. In \cite{jian2013regulated}, household load variance is minimized by EV control. Tap changer effects on peak shaving from EV was studied in \cite{alam2015controllable}. Price-based programs of demand response, which is based on dynamic pricing rates, can be employed to flatten the demand curve by offering a higher price during peak periods and a lower price during off-peak periods \cite{albadi2008summary}. Household appliances can also participate in peak shaving as shown in \cite{caprino2014peak}. Meanwhile, power at the PCC can be limited to a certain range during cost-based \cite{malysz2014optimal,worthmann2015distributed} or risk-based \cite{khodabakhsh2016optimal} scheduling.\par

Most of the aforementioned research relies on a cost-optimization framework, which highly depends on the specific electricity price profile or market mechanism. Consequently, it is hard to clarify the optimal solution of load variance minimization constrained only by physical limits. A recent theoretical study on peak shaving in \cite{levron2012power} proved the optimal solution under limited storage capacity. It was shown that the infinity norm of the power at PCC is minimized if the energy at PCC takes the shortest path within the energy band between the load energy and the energy summation of load and storage limit. Such a theoretical study helps clarify the physical limits under correct mathematical principle without mixed effects of cost and other considerations. For now, there is no theoretical formulation for load variance minimization. \par

Thus, this paper aims at theoretical study on load variance minimization. The infinity norm minimization-based battery energy storage (BES) scheduling is formulated to optimally flatten the load. This formulation is more general than the shortest path principle \cite{levron2012power}, which is only applicable to peak shaving. Moreover, an equivalent formulation is proposed to convert the optimization problem into a linear program. The radial distribution network power flow (DistFlow) proposed in \cite{baran1989network,baran1989sizing} is employed in this study with several simplifications made based on \cite{yeh2012adaptive}. The impacts of capacity limit, voltage constraints and location of BES units on load variance minimization performance are investigated.\par
The remainder of this paper is organized as follows. Section \ref{sec_2} introduces the infinity norm minimization-based scheduling formulation and its equivalent form. The BES model and radial distribution network power flow are introduced as well. Section \ref{sec_3} presents the case study on a lumped system as well as a radial feeder. Finally, the conclusion and future work are discussed in Section \ref{sec_4}.

\section{PROBLEM FORMULATION}\label{sec_2}
\subsection{Objective Function}
\begin{figure}[t]
	\centering
	\includegraphics[width=3.5 in]{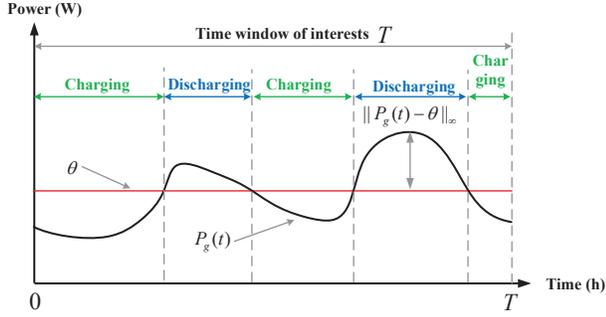}
	\vspace{-0.15in}
	\caption{Load flattening objective.}
	\label{fig_LF_OB}
\end{figure}

Let $P_{Bc}(t)$ and $P_{Bd}(t)$ denote total BES charging and discharging powers, respectively. Let $P_{L}(t)$ denote the forecasted total load profile, which is assumed to be known in this paper. Loss is neglected for now. Then, the power at PCC $P_{g}(t)$ can be expressed as follows
\begin{equation}
\begin{aligned}
P_{g}(t)=P_{Bc}(t)- P_{Bd}(t)+P_{L}(t)
\end{aligned}
\end{equation}
where $P_{Bc}(t)>0$ and $P_{Bd}(t)>0$. The objective is to minimize the demand peak-valley gap. To explore the optimality, the ideal case is set as the target curve, where there is no peak-valley gap illustrated as the red line in Fig. \ref{fig_LF_OB}. Then, the objective can be expressed as the infinity norm minimization as follows
\begin{equation}\label{eq_inf_min}
\begin{aligned}
&\underset{P_{Bc}(t),P_{Bd}(t)}{\min}\lVert P_{Bc}(t)- P_{Bd}(t)+P_{L}(t)-\theta\rVert_{\infty}\\
\end{aligned}
\end{equation}
In robust control, instead of minimizing $||T_{zw}||_{\infty}$, it is usually desired to minimize the upper bound $\gamma$, where $||T_{zw}||_{\infty}<\gamma$ and $\gamma>0$ \cite{mahmoud1999h}. Similar idea is applied into the objective reformulation as the objective in (\ref{eq_inf_min}) is equivalent to
\begin{equation}
\begin{aligned}
\underset{P_{Bc}(t),P_{Bd}(t)}{\min}\quad K\\
\end{aligned}
\end{equation}
such that
\begin{equation}\label{eq_main1_cons}
\begin{aligned}
 0\leq\mid P_{Bc}(t)-P_{Bd}(t)+P_{L}(t)-\theta\mid\leq K
\end{aligned}
\end{equation}\par
There are two ways to deal with the target curve $\theta$. One is to treat $\theta$ as a given constant. The second approach is to regard $\theta$ as a decision variable. Since here the objective is to minimize the variance of the power at PCC, so $\theta$ needs to be limited between the valley and peak of forecasted load. Then the alternative problem can be formulated as
\begin{equation}
\begin{aligned}
&\underset{P_{Bc}(t),P_{Bd}(t),\theta}{\min}\quad \alpha K+\beta\theta
\end{aligned}
\end{equation}
such that
\begin{equation}\label{eq_main2_cons}
\begin{aligned}
& 0\leq\mid P_{Bc}(t)-P_{Bd}(t)+P_{L}(t)-\theta\mid\leq K\\
& \min{P_{L}(t)}\leq\theta\leq\max{P_{L}(t)},\alpha>0,\beta>0
\end{aligned}
\end{equation}\par

\begin{remark}
	The formulation in \cite{levron2012power} employs the infinity norm in the following form
	\begin{equation}
	\begin{aligned}
	||z(t)||_{\infty}=\lim_{m\rightarrow\infty}(\int\limits_{0}^{T}|z(\tau)|^{m}\text{d}\tau)^{1/m}\\
	\end{aligned}
	\end{equation}
	The shortest path principle can be applied when $|z(\tau)|^{m}$ is monotonically increase. It is true when performing peak shaving without reverse load flow to main grid since $z(\tau)=P_{g}(\tau)>0$. But it does not hold when minimizing load variance, where $z(\tau)=P_{g}(\tau)-\theta$ is not always positive. As seen, the formulation proposed above is more general.
\end{remark}

\subsection{Battery Energy Storage Model}
Let $\mathcal{E}$ denote the set of buses that have BES and load connected. The battery model can be expressed as follows \cite{kou2017distributed}
\begin{equation}\label{eq_bes_dyn}
\begin{aligned}
& E_{b,i}(t)=E_{b,i}(t-1)+\eta T P_{bc,i}(t)-\frac{1}{\eta}T P_{bd,i}(t)\quad\forall i\in \mathcal{E}\\
\end{aligned}
\end{equation}
where $P_{bc,i}(t)$ and $P_{bd,i}(t)$ are charging and discharging power of BES at bus $i$ in kW. $E_{b,i}(t)$ is the available capacity of BES at bus $i$ (kW$\cdot$h). $\eta$ represents the BES charging/discharging efficiency. $T$ is the time interval. The BES capacity limit can be expressed by the following constraints
\begin{equation}\label{eq_bes_cap}
\begin{aligned}
&\text{SOC}_{\min}B_{\text{cap}}\leq E_{b,i}(t)\leq\text{SOC}_{\max}B_{\text{cap}}\quad\forall i\in \mathcal{E}
\end{aligned}
\end{equation}
where $B_{\text{cap}}$ is the rated capacity in kW$\cdot$h. $\text{SOC}_{\min}$ and $\text{SOC}_{\max}$ are the state-of-charge limits in percentage. In general, simultaneous charging and discharging is unrealistic. Binary variable can be introduced to force only one action to be activated during each scheduling interval $T$ \cite{liu2017microgrid} as follows
\begin{equation}\label{eq_bes_power}
\begin{aligned}
& P_{Bc}(t)=\sum_{i\in \mathcal{E}}P_{bc,i}(t),P_{Bd}(t)=\sum_{i\in \mathcal{E}}P_{bd,i}(t)\\
& 0\leq P_{bc,i}(t)\leq P_{bc,\max}m_{bc,i}(t)        \quad\forall i\in \mathcal{E}\\
& 0\leq P_{bd,i}(t)\leq P_{bd,\max}m_{bd,i}(t)       \quad\forall i\in \mathcal{E}\\
& m_{bc,i}(t)+m_{bd,i}(t)\leq 1  \quad\forall i\in \mathcal{E}
\end{aligned}
\end{equation}
where $P_{bc,\max}$ and $P_{bd,\max}$ are the BES charging and discharging power limits. $m_{bc,i}(t)$ and $m_{bd,i}(t)$ are binary variables representing the operating modes of charging and discharging. 1 means the corresponding mode is activated while 0 stands for deactivation. The last constraint in Eq. (\ref{eq_bes_power}) ensures no simultaneous charging and discharging situation.

\subsection{Radial Distribution Network Power Flow}
\begin{figure}[t]
	\centering
	\includegraphics[width=3 in]{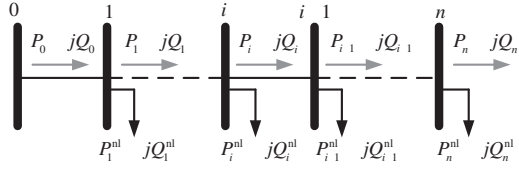}
	\vspace{-0.15in}
	\caption{One line diagram of a main distribution feeder.}
	\label{fig_RadialNetwork}
\end{figure}

The distribution network power flow (DistFlow) is borrowed from \cite{baran1989network} and more generally from \cite{baran1989sizing} including lateral branches, which has been employed in \cite{wang2015coordinated} for microgrid scheduling. 
Consider a radial distribution network shown in Fig. \ref{fig_RadialNetwork}. Let $\mathcal{N}$ denote the set of buses that only have load connected. The following equations can be used to describe the complex power flows at each node $i$ at time $t$
\begin{equation}
\begin{aligned}
& P_{i+1}(t)=P_{i}(t) - r_{i}\dfrac{P_{i}^{2}(t)+Q_{i}^{2}(t)}{V_{i}^{2}(t)} - P_{\text{nl},i+1}(t)\\
& Q_{i+1}(t)=Q_{i}(t) - x_{i}\dfrac{P_{i}^{2}(t)+Q_{i}^{2}(t)}{V_{i}^{2}(t)} - Q_{\text{nl},i+1}(t)\\
& V_{i+1}^{2}(t)=V_{i}^{2}(t) - 2[r_{i}P_{i}(t)+x_{i}Q_{i}(t)] + (r_{i}^2+x_{i}^{2})\dfrac{P_{i}^2(t)+Q_{i}^2(t)}{V_{i}^2(t)}
\end{aligned}
\end{equation}
where nl stands for net load. $x_{i}$ is inductance from branch $i$ to $i+1$. $P_{i}$ and $Q_{i}$ are the load flow from branch $i$ to $i+1$. The first approximation is made by dropping the quadratic terms as the branch losses are much smaller than the branch power \cite{baran1989sizing}. Then, the simplified DistFlow is shown as
\begin{equation}
\begin{aligned}
& P_{i+1}(t)=P_{i}(t)-P_{\text{nl},i+1}(t)\\
& Q_{i+1}(t)=Q_{i}(t)-Q_{\text{nl},i+1}(t)\\
& V_{i+1}^{2}(t)=V_{i}^{2}(t) - 2[r_{i}P_{i}(t)+x_{i}Q_{i}(t)]
\end{aligned}
\end{equation}
The second approximation is made by the following assumption in \cite{yeh2012adaptive}
\begin{equation}
\begin{aligned}
\left[ V_{i}(t)-V_{0}(t)\right]  ^{2}\approx 0,
\end{aligned}
\end{equation}
which leads to
\begin{equation}
\begin{aligned}
V_{i}^2(t)\approx V_{0}^{2}(t) + 2V_{0}(t)[V_{i}(t)-V_{0}(t)]
\end{aligned}
\end{equation}
The rationality of the second approximation lies in the fact that the per unit voltage variation along the line remain within the bounds for proper operation of a distribution system as
\begin{equation}\label{eq_volt_cons}
\begin{aligned}
1-\epsilon \leq V_{i}(t) \leq 1+\epsilon
\end{aligned}
\end{equation}
where $\epsilon > 0$ is generally very small \cite{yeh2012adaptive}. After the two approximations, the DistFlow equations become
\begin{equation}\label{eq_simp_flow}
\begin{aligned}
& P_{i+1}(t)=P_{i}(t)-P_{i+1}^{\text{nl}}(t),Q_{i+1}(t)=Q_{i}(t)-Q_{i+1}^{\text{nl}}(t)\\
& V_{i+1}(t)=V_{i}(t)-\dfrac{r_{i}P_{i}(t)+x_{i}Q_{i}(t)}{V_{0}(t)},P_{L}(t)=\sum_{i\in \mathcal{E}\cup\mathcal{N}}P_{l,i}(t)\\
& P_{\text{nl},i}(t)=P_{l,i}(t),Q_{\text{nl},i}(t)=Q_{l,i}(t)\quad\forall i\in \mathcal{N}\\
& P_{\text{nl},i}(t)=P_{l,i}(t)+P_{bc,i}(t)-P_{bd,i}(t)\quad\forall i\in \mathcal{E}\\
& Q_{\text{nl},i}(t)=Q_{l,i}(t)+Q_{bc,i}(t)-Q_{bd,i}(t)\quad\forall i\in \mathcal{E}
\end{aligned}
\end{equation}
The terminal condition is given as $P_{n}(t)=0$ and $Q_{n}(t)=0$ for all $t$. As we can see, by restricting the voltage variation the BES power is confined. Consequently, this turns out to be another constraint for load variance minimization.\par

\section{CASE STUDY}\label{sec_3}
In this section, we will evaluate the performance of the proposed technique on a 12 kV distribution system. The study is performed on two versions of the system, i.e., a lumped version and a radial version, for different purposes. 
The parameters are given as $P_{\text{base}}=1\text{ MW}, \varepsilon=0.05,\text{SOC}_{\min}=5\%, \text{SOC}_{\max}=95\%, \eta=0.9,T=1[\text{h}], E_{b,i}(0)=\text{SOC}_{\min}\times B_{\text{cap}}\text{ for }i\in\mathcal{E},V_{0}(t)=1.02\text{ for any }t$. $P_{bc,\max}$ and $P_{bd,\max}$ are assumed to be sufficiently large.

\subsection{A Lumped System without Voltage Constraint}
\begin{figure}[b]
	\centering
	\includegraphics[width=2.2 in]{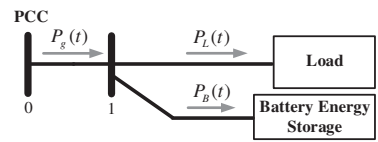}
	\vspace{-0.15in}
	\caption{A lumped system.}
	\label{fig_LumpedSystem}
\end{figure}
\begin{figure*}[t]
	\centering
	\includegraphics[width=5.9 in]{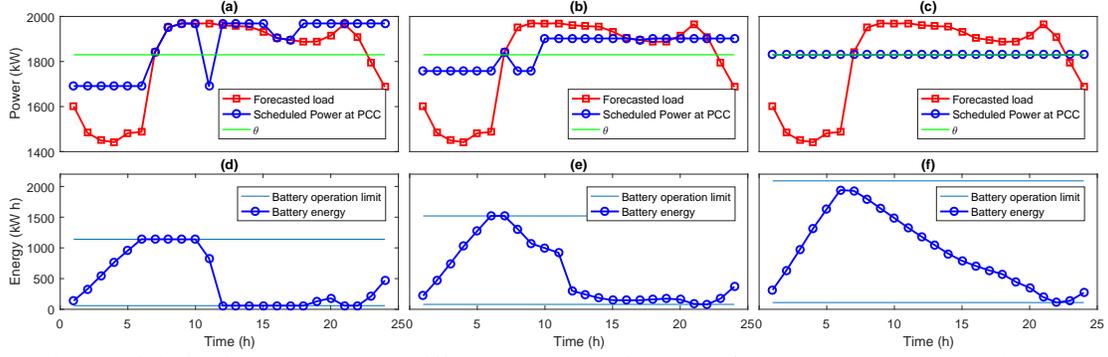}
	\vspace{-0.15in}
	\caption{Load variance minimization results under different BES capacity when $\theta=1830$: (a), (b), (c) and (d), (e), (f) are power curves and BES energy curves when $B_{\text{cap}}$ equals to 1200 kW$\cdot$h, 1600 kW$\cdot$h and 2200 kW$\cdot$h, respectively. }
	\label{fig_Diff_cap}
\end{figure*}

The lumped system shown in Fig. \ref{fig_LumpedSystem} is employed to study the property of the formulation. The system consists of one aggregated BES and load, and is assumed to be lossless. Consider the first optimization problem
\begin{equation}\label{eq_OP1}
\begin{aligned}
&\underset{P_{Bc}(t),P_{Bd}(t)}{\min}\quad K\\
&\text{s.t. }(\ref{eq_main1_cons}),(\ref{eq_bes_dyn}),(\ref{eq_bes_cap}),(\ref{eq_bes_power})
\end{aligned}
\end{equation}
The problem in (\ref{eq_OP1}) is a mix-integer linear programming (MILP). Let $\theta=1830$. The load flattening results when BES capacity equals to 1200 kW$\cdot$h, 1600 kW$\cdot$h and 2200 kW$\cdot$h are shown in Fig. \ref{fig_Diff_cap}. Based on the scheduling results using (\ref{eq_OP1}), when $B_{\text{cap}}\geq 2040 \text{ kW}\cdot\text{h}$, the load flattening attains the best performance, where $K=0$.
\par

\begin{figure}[t]
	\centering
	\includegraphics[width=2.7 in]{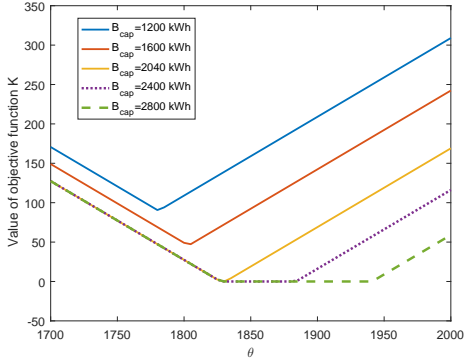}
	\vspace{-0.1in}
	\caption{Relationship between value of objective function and choice of $\theta$ under different BES capacity limits. }
	\label{fig_theta_K}
\end{figure}

As expected, the value of objective function varies with the choice of $\theta$. The relationship curve under different BES capacity is studied and illustrated in Fig. \ref{fig_theta_K}. As shown, when the BES capacity is less than or equal to the critical capacity limit, there is only one choice of $\theta$ that can reach the best load flattening performance under that particular capacity limit. After the critical limit, larger BES capacity will lead to wider range of $\theta$ that can minimize the load variance to zero shown as the flat bottom in Fig. \ref{fig_theta_K}. In these scenarios, it is always desired to use the smallest $\theta$ for cost reduction, which corresponds to the left turning point at the flat bottom. This critical choice of $\theta$ enables the energy balance between peak shaving and valley filling. It should be mentioned that the curves will be altered when the BES unit starts from a different initial condition.

\begin{figure}[t]
	\centering
	\includegraphics[width=2 in]{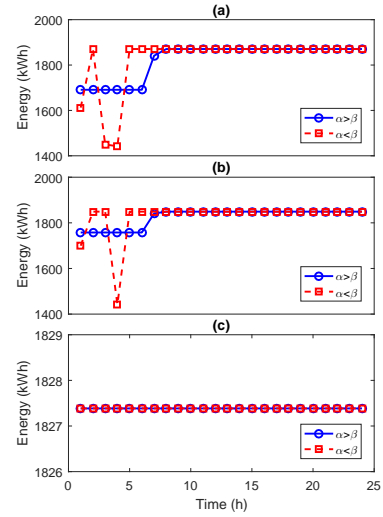}
	\vspace{-0.1in}
	\caption{Scheduled power at PCC under different objective weighting factors and BES capacity. (a) 1200 kW$\cdot$h, (b) 1600 kW$\cdot$h and (c) 2040 kW$\cdot$h.}
	\label{fig_AB_com}
\end{figure}

Regarding $\theta $ as another decision variable yields:
\begin{equation}\label{eq_OP2}
\begin{aligned}
&\underset{P_{Bc}(t),P_{Bd}(t)}{\min}\quad \alpha K+\beta\theta\\
&\text{s.t. }(\ref{eq_main2_cons}),(\ref{eq_bes_dyn}),(\ref{eq_bes_cap}),(\ref{eq_bes_power})
\end{aligned}
\end{equation}
The scheduling results are summarized in Tab. \ref{tab_op2}. When $\alpha>\beta$, the values of the objective function are exactly those turning points in Fig. \ref{fig_theta_K}. However, once $\alpha<\beta$, $\theta$ is minimized with priority and pushed to its lower limit. Thus, the infinity norms are much larger than those in the first case. The scheduled power at PCC of these two cases with different BES capacity is shown in Fig. (\ref{fig_AB_com}). When the BES capacity is larger than the critical value, although the values of the objective function are different, the scheduled results are the same. But it is not the case when the BES capacity is not adequate for a perfect flattening. \par
\begin{table*}[htb]
	\caption{Results of Problem in (\ref{eq_OP2})} \label{tab_op2} 
	\centering		
	\begin{tabular}{lclclclclclcl}
		\hline
		Capacity            & 1200 kWh              & 1600 kWh        & 2040 kWh        & 2400 kWh        &  2800 kWh \\ \hline
		$\alpha>\beta$   & $K=89.8$,$\theta=1781$   & $K=45.7$,$\theta=1803$  & $K=0$,  $\theta=1827$  &  $K=0$,$\theta=1827$ & $K=0$,$\theta=1827$  \\ \hline
		$\alpha<\beta$   & $K=428.7$,$\theta=1442$  & $K=407.1$,$\theta=1442$ & $K=385.4$,$\theta=1442$  &  $K=385.3$,$\theta=1442$ & $K=385.3$,$\theta=1442$ 	   \\ \hline
	\end{tabular}
\end{table*}
\par

\subsection{A Radial Feeder with Voltage Constraints}
\begin{figure}[t]
	\centering
	\includegraphics[width=2.7 in]{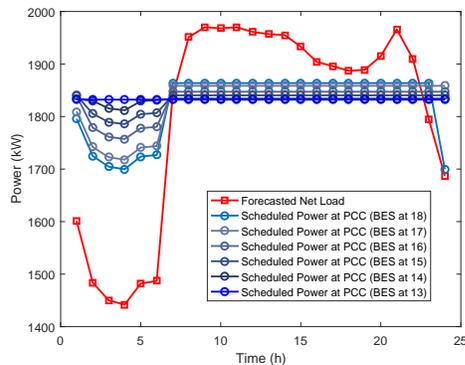}
	\vspace{-0.1in}
	\caption{Scheduled power at PCC when one BES with capacity of 2200 kW$\cdot$h is deployed at different buses.}
	\label{fig_BES_location}
\end{figure}

In this section, the lumped system is expanded into a 18-bus radial feeder to study the voltage constraint. Loss is considered in this case. The same total net active load in previous section is used, and distributed to each bus according to Tab. \ref{tab_feeder}. The reactive load is assumed to be fixed and given in Tab. \ref{tab_feeder} as well. Based on the study in the last subsection, Formulation 2 is chosen with larger weighting factor on infinity norm $K$ than target $\theta$, which leads to
\begin{equation}\label{eq_OP3}
\begin{aligned}
&\underset{P_{Bc}(t),P_{Bd}(t)}{\min}\quad \alpha K+\beta\theta\\
&\text{s.t. }(\ref{eq_main2_cons}),(\ref{eq_bes_dyn}),(\ref{eq_bes_cap}),(\ref{eq_bes_power}),(\ref{eq_volt_cons}),(\ref{eq_simp_flow}),\alpha>\beta
\end{aligned}
\end{equation}
\begin{table}[htb]
	\caption{Radial Feeder Data}\label{tab_feeder} 
	\centering		
	\begin{tabular}{lclclclclclcl}
		\hline
		      &      &             &  Loads on to-node &\\
		From  & To   & R (p.u.)  & X (p.u.)    &$\%$ in Total P   & Q in p.u. \\ \hline
		  0 &   1 &  0.000574  &0.000293 &  6.64 & 0.06 \\ \hline
		  1 &   2 &  0.00307   &0.001564 &  5.98 & 0.04 \\ \hline
		  2 &   3 &  0.002279    &0.001161 &  7.97 & 0.08 \\ \hline
		  3 &   4 &  0.002373  &0.001209 &  3.99 & 0.03 \\ \hline
		  4 &   5 &  0.0051  &0.004402 &  3.99 & 0.02 \\ \hline
		  5 &   6 &  0.001166  &0.003853 &  13.29 & 0.1 \\ \hline
		  6 &   7 &  0.00443  &0.001464 &  13.29 & 0.1 \\ \hline
		  7 &   8 &  0.006413  &0.004608 &  3.99 & 0.02 \\ \hline
		  8 &   9 &  0.006501  &0.004608 &  3.99 & 0.02 \\ \hline
		  9 &   10 & 0.001224  &0.000405 & 2.99 & 0.03 \\ \hline
		  10 &   11 & 0.002331  &0.000771 &  3.99 & 0.035 \\ \hline
		  11 &   12 & 0.009141  &0.007192 &  3.99 & 0.035 \\ \hline
		  12 &   13 & 0.003372  &0.004439 &  7.97 & 0.08 \\ \hline
		  13 &   14 & 0.00368  &0.003275 &  3.99 & 0.01 \\ \hline
		  14 &   15 & 0.004647 &0.003394 &  3.99 & 0.02 \\ \hline
		  15 &   16 & 0.008026  &0.010716 &  3.99 & 0.02 \\ \hline
		  16 &   17 & 0.004558   &0.003574 &  5.98 & 0.04 \\ \hline
	\end{tabular}
\end{table}
One BES with adequate size of 2200 kW$\cdot$h is deployed in different nodes and the scheduled power at PCC is shown in Fig. \ref{fig_BES_location}. Due to the voltage constraint, when the BES is deployed at the end of the network, the charging power is limited to further limit the voltage drop along the line. When the deployment location is moving forward, the scheduled power is approaching the best performance as expected. The best performance is achieved when the BES is deployed at bus 13 or any bus ahead of bus 13.
\par

\section{CONCLUSIONS AND FUTURE WORKS}\label{sec_4}
This paper proposes an infinity norm minimization problem aiming at minimizing the load variance. Two linear programs are formulated to solve the proposed problem. The equivalence of these two problems are observed under certain conditions. A case study is performed with capacity-limited battery energy storage model and the simplified power flow model of a radial distribution network. The critical capacity of BES is obtained for the best performance, meaning zero variance of the scheduled power at PCC. Deployment location is also studied. It is shown that for better scheduling performance it is desired to deploy the BES units in the front of a radial network due to the voltage constraints.\par

Based on the proposed formulation, optimal allocation and sizing of BES units for load variance minimization can be determined. Different types of DER like fuel cell \cite{raoufat2016performance} can be integrated and analyzed. Since the formulation provides capability of reshaping power at PCC, a resiliency-oriented or a market-oriented scheduling problem can be considered. Distributed coordination of BES units and tap changers over a large distribution network with more sophisticated power flow model would be an interesting extension. 

\bibliographystyle{IEEEtran}
\bibliography{IEEEabrv_zyc,Ref_LF}  

\begin{thebibliography}{10}
\providecommand{\url}[1]{#1}
\csname url@samestyle\endcsname
\providecommand{\newblock}{\relax}
\providecommand{\bibinfo}[2]{#2}
\providecommand{\BIBentrySTDinterwordspacing}{\spaceskip=0pt\relax}
\providecommand{\BIBentryALTinterwordstretchfactor}{4}
\providecommand{\BIBentryALTinterwordspacing}{\spaceskip=\fontdimen2\font plus
\BIBentryALTinterwordstretchfactor\fontdimen3\font minus
  \fontdimen4\font\relax}
\providecommand{\BIBforeignlanguage}[2]{{%
\expandafter\ifx\csname l@#1\endcsname\relax
\typeout{** WARNING: IEEEtran.bst: No hyphenation pattern has been}%
\typeout{** loaded for the language `#1'. Using the pattern for}%
\typeout{** the default language instead.}%
\else
\language=\csname l@#1\endcsname
\fi
#2}}
\providecommand{\BIBdecl}{\relax}
\BIBdecl

\bibitem{duck_curve}
\BIBentryALTinterwordspacing
What the duck curve tells us about managing a green grid. California ISO.
  [Online]. Available:
  \url{https://www.caiso.com/Documents/FlexibleResourcesHelpRenewables_FastFacts.pdf}
\BIBentrySTDinterwordspacing

\bibitem{sanandaji2016ramping}
B.~M. Sanandaji, T.~L. Vincent, and K.~Poolla, ``Ramping rate flexibility of
  residential hvac loads,'' \emph{{IEEE} Trans. Sustain. Energy}, vol.~7,
  no.~2, pp. 865--874, 2016.

\bibitem{majzoobi2016application}
A.~Majzoobi and A.~Khodaei, ``Application of microgrids in supporting
  distribution grid flexibility,'' \emph{{IEEE} Trans. Power Syst.}, vol.~32,
  no.~5, pp. 3660--3669, 2017.

\bibitem{wang2013grid}
Z.~Wang and S.~Wang, ``Grid power peak shaving and valley filling using
  vehicle-to-grid systems,'' \emph{{IEEE} Trans. Power Del.}, vol.~28, no.~3,
  pp. 1822--1829, 2013.

\bibitem{rowe2014peak}
M.~Rowe, T.~Yunusov, S.~Haben, C.~Singleton, W.~Holderbaum, and B.~Potter, ``A
  peak reduction scheduling algorithm for storage devices on the low voltage
  network,'' \emph{{IEEE} Trans. Smart Grid}, vol.~5, no.~4, pp. 2115--2124,
  2014.

\bibitem{joshi2015day}
K.~A. Joshi and N.~M. Pindoriya, ``Day-ahead dispatch of battery energy storage
  system for peak load shaving and load leveling in low voltage unbalance
  distribution networks,'' in \emph{Proc. {IEEE} {PES} Innov. Smart Grid
  Technol. (ISGT)}, 2015, pp. 1--5.

\bibitem{jian2013regulated}
L.~Jian, H.~Xue, G.~Xu, X.~Zhu, D.~Zhao, and Z.~Shao, ``Regulated charging of
  plug-in hybrid electric vehicles for minimizing load variance in household
  smart microgrid,'' \emph{{IEEE} Trans. Ind. Electron.}, vol.~60, no.~8, pp.
  3218--3226, 2013.

\bibitem{alam2015controllable}
M.~J.~E. Alam, K.~M. Muttaqi, and D.~Sutanto, ``A controllable local
  peak-shaving strategy for effective utilization of pev battery capacity for
  distribution network support,'' \emph{{IEEE} Trans. Ind. Appl.}, vol.~51,
  no.~3, pp. 2030--2037, 2015.

\bibitem{albadi2008summary}
M.~H. Albadi and E.~F. El-Saadany, ``A summary of demand response in
  electricity markets,'' \emph{Electr. Power Syst. Res.}, vol.~78, no.~11, pp.
  1989--1996, 2008.

\bibitem{caprino2014peak}
D.~Caprino, M.~L. Della~Vedova, and T.~Facchinetti, ``Peak shaving through
  real-time scheduling of household appliances,'' \emph{Energy and Buildings},
  vol.~75, pp. 133--148, 2014.

\bibitem{malysz2014optimal}
P.~Malysz, S.~Sirouspour, and A.~Emadi, ``An optimal energy storage control
  strategy for grid-connected microgrids,'' \emph{{IEEE} Trans. Smart Grid},
  vol.~5, no.~4, pp. 1785--1796, 2014.

\bibitem{worthmann2015distributed}
K.~Worthmann, C.~M. Kellett, P.~Braun, L.~Gr{\"u}ne, and S.~R. Weller,
  ``Distributed and decentralized control of residential energy systems
  incorporating battery storage,'' \emph{{IEEE} Trans. Smart Grid}, vol.~6,
  no.~4, pp. 1914--1923, 2015.

\bibitem{khodabakhsh2016optimal}
R.~Khodabakhsh and S.~Sirouspour, ``Optimal control of energy storage in a
  microgrid by minimizing conditional value-at-risk,'' \emph{{IEEE} Trans.
  Sustain. Energy}, vol.~7, no.~3, pp. 1264--1273, 2016.

\bibitem{levron2012power}
Y.~Levron and D.~Shmilovitz, ``Power systems’ optimal peak-shaving applying
  secondary storage,'' \emph{Electr. Power Syst. Res.}, vol.~89, pp. 80--84,
  2012.

\bibitem{baran1989network}
M.~E. Baran and F.~F. Wu, ``Network reconfiguration in distribution systems for
  loss reduction and load balancing,'' \emph{{IEEE} Trans. Power Del.}, vol.~4,
  no.~2, pp. 1401--1407, 1989.

\bibitem{baran1989sizing}
M.~Baran and F.~F. Wu, ``Optimal sizing of capacitors placed on a radial
  distribution system,'' \emph{{IEEE} Trans. Power Del.}, vol.~4, no.~1, pp.
  735--743, 1989.

\bibitem{yeh2012adaptive}
H.-G. Yeh, D.~F. Gayme, and S.~H. Low, ``Adaptive var control for distribution
  circuits with photovoltaic generators,'' \emph{{IEEE} Trans. Power Syst.},
  vol.~27, no.~3, pp. 1656--1663, 2012.

\bibitem{mahmoud1999h}
M.~Mahmoud and M.~Zribi, ``H-infinity-controllers for time-delay systems using
  linear matrix inequalities,'' \emph{J. Optimiz. Theory and Applic.}, vol.
  100, no.~1, pp. 89--122, 1999.

\bibitem{kou2017distributed}
P.~Kou, D.~Liang, and L.~Gao, ``Distributed empc of multiple microgrids for
  coordinated stochastic energy management,'' \emph{Applied Energy}, vol. 185,
  pp. 939--952, 2017.

\bibitem{liu2017microgrid}
G.~Liu, M.~Starke, B.~Xiao, X.~Zhang, and K.~Tomsovic, ``Microgrid optimal
  scheduling with chance-constrained islanding capability,'' \emph{Electr.
  Power Syst. Res.}, vol. 145, pp. 197--206, 2017.

\bibitem{wang2015coordinated}
Z.~Wang, B.~Chen, J.~Wang, M.~M. Begovic, and C.~Chen, ``Coordinated energy
  management of networked microgrids in distribution systems,'' \emph{{IEEE}
  Trans. Smart Grid}, vol.~6, no.~1, pp. 45--53, 2015.

\bibitem{raoufat2016performance}
M.~E. Raoufat, A.~Khayatian, and A.~Mojallal, ``Performance recovery of voltage
  source converters with application to grid-connected fuel cell dgs,''
  \emph{{IEEE} Trans. Smart Grid}, 2016.

\end{thebibliography}

\end{document}